\documentstyle[12pt,aasms4]{article}

\def\Ms{M_\odot}

\def\rs{\rm s}
\def\rs1{\rm s^{-1}}

\def\rcm{\rm cm}
\def\rcm2{\rm cm^{-2}}

\def\c2r{\chi^2_\nu}
\def\zeta{A_{Fe}}

\def\etal{{\it et al. }}
\def\aas{{\it A\&AS}}
\begin{document}

\title{The X-ray afterglow 
of the Gamma-ray burst of May 8, 1997: spectral variability and
possible evidence of an iron line}

\author{
L. Piro \altaffilmark{1},
E. Costa \altaffilmark{1}, 
M. Feroci \altaffilmark{1},
 F. Frontera \altaffilmark{2}\altaffilmark{3}, 
L. Amati \altaffilmark{2},
D. Dal Fiume\altaffilmark{2},
L. A. Antonelli \altaffilmark{4},
J. Heise\altaffilmark{5},
 J. in 't Zand\altaffilmark{5},
A. Owens\altaffilmark{6},
A.N. Parmar\altaffilmark{6},
G. Cusumano\altaffilmark{7},
M. Vietri\altaffilmark{8},
G.C. Perola \altaffilmark{8}
}
\altaffiltext{1}{Istituto Astrofisica Spaziale, C.N.R.,
Via Fosso del Cavaliere 100, 00133 Rome}
\altaffiltext{2}{ Istituto TeSRE, C.N.R.,
Via Gobetti 101, 40129 Bologna, Italy}
\altaffiltext{3}{Dip. Fisica, Universita' Ferrara, Via Paradiso 12, 
Ferrara, Italy}
\altaffiltext{4}{BeppoSAX Scientific Data Center;
present address: Osservatorio Astronomico di Roma}
\altaffiltext{5}{ Space Research Organization in the 
Netherlands, Sorbonnelaan 2, 3584 CA Utrecht, The Netherlands}
\altaffiltext{6}{Space Science Department of ESA, ESTEC,
2200 AG Noordwijk, The Netherlands}
\altaffiltext{7}{Istituto Fisica Cosmica e Appl. Calc. Informatico, C.N.R.,
Via La Malfa 153, 90138 Palermo, Italy}
\altaffiltext{8}{Dipartimento di Fisica, Universita' Roma Tre,
Via della Vasca Navale 84, 00146 Roma, Italy}

\begin{abstract}
We report the possible detection (99.3\% of statistical
significance) of
redshifted Fe iron line emission 
in the X-ray afterglow of 
Gamma-ray burst GRB970508 observed by BeppoSAX. 
Its energy is 
consistent with the redshift of the putative host
galaxy  determined from optical spectroscopy.
The line disappeared  $\sim 1\ $ day after the burst.
We have also analyzed the spectral variability 
during the outburst event that characterizes the X-ray afterglow
of this GRB. The spectrum gets harder during the flare, turning
to steep when the flux decreases.
The variability, intensity and width of the line indicate that
the emitting region should have a mass 
$\gtrsim 0.5\Ms$ (assuming the iron abundance 
similar to its solar value),
 a size of
$\sim 3\times 10^{15} cm$, is distributed anisotropically,
and is moving with sub-relativistic speed.
In contrast to the fairly clean environment
expected in the merging of two neutron stars,
the observed line properties would imply that
the site of the burst is  embedded in a large
mass of material, consistent with pre-explosion ejecta
of a very massive star.
This material could  be related with the outburst observed
in the afterglow 1 day after the GRB and with 
the spectral variations measured
during this phase. 
\end{abstract}


\section{Introduction}
Distance - scale determination of Gamma-ray bursts (GRB) has been 
one of the most important achievements of 
astrophysics in recent years. Accurate and fast localization of the 
prompt and afterglow emission  bursts (i.e.\cite{Costa97}) by the
Italian/Dutch X--ray satellite BeppoSAX (\cite{P95,B97}) led to the 
identification of optical counterparts (\cite{vpd97})
and ultimately to spectral 
measurements of a redshift (\cite{metzger}). 
While the extragalactic origin of GRB has gathered  solid 
evidence in its 
support, the source of the large  energy implied by
their distance
 is still speculative.

Direct  information on the nature  
of the central engine of GRB  can be 
derived by studying the nearby 
environment; for example using line 
spectroscopy. The star forming 
region and massive pre-explosion winds 
associated with the hypernova scenario
(\cite{pacz98})
 imply a mass rich environment down to short 
($\sim 10^{15} cm$) distance scales.
In contrast, NS-NS merging should happen
in a fairly clean environment, because such objects are
expected to form with significant speeds which can lead them 
at least a few parsecs away from the regions of
star formation in which they formed. 
Current line measurements are inconclusive, because all the 
spectral features  observed so far are signatures
 produced by the host galaxy rather 
than at the burst site. GRB980425 is an 
exception, but its tantalizing 
association with SN 1998bw 
(\cite{galama98,k98b}) requires clarification.

The measurement of X-ray lines emitted directly by the GRB or its afterglow
could provide  a direct measurement of the distance and
probe into the nature of the central environment (\cite{pl98}). 
 Fe K line emission is very promising in this respect. Theoretical
computations have been carried out by several authors in the framework
of GRB (\cite{mr98,bdcl98,ghi98}). Most of this work deals with
features generated within the interstellar medium (ISM) of the host 
galaxy, i.e.
on scale larger than several parsecs. However, even in the favourable case of
dense regions of stellar formation (e.g. the site of GRB formation in the
hypernova scenario), iron emission lines 
would be hard to
detect with  current X-ray instrumentation.
Lately, M\'esz\'aros and Rees (1998)  have shown that the
circumburst environment created by the stellar wind before the
explosion of the hypernova could yield a line of substantial
intensity. A similarly favourable situation should be expected in
a variation of the hypernova scenario, -- the SupraNova (\cite{vs98}),
where the GRB is shortly preceded by a supernova explosion with the ejection
of an iron--rich  massive shell.
In these cases, and for a isotropic distribuition 
of material around the source, 
a tight upper limit on the column density $N_H<10^{24}$cm$^{-2}$
derives from the requirement that the Thomson optical
depth be less than a few, to avoid smearing out the
the short time-scale structure
of the GRB by Thomson scattering (\cite{bdcl98}).
Furthermore, Fe absorption features of 
stregth comparable to that in emission should also be detected.

It is also conceivable that the impact of the  
 relativistic shell that
produced the original GRB  on these ejecta
could provide an additional 
energy input in the afterglow.

GRB970508 is the most promising candidate of the
BeppoSAX afterglows for the search of line emission.
 It is characterized by a large outbursting event 
during its afterglow phase (\cite{Piro98}) and has
the highest signal to noise ratio of the BeppoSAX GRB afterglows. 
In this Letter we present results of a spectral analysis
of the afterglow of GRB970508 and discuss their implications
on the nature of the central engine of GRB.

\section{Observations  of the X-ray afterglow spectrum of GB970508}

The Gamma-ray burst GRB970508 was detected
by the Gamma-ray Burst Monitor (GRBM)
aboard BeppoSAX (\cite{Piro98}). The Wide Field Camera (WFC)
 provided a
few arcmin position acquired with the Narrow Field Instruments 6 hours
after the burst, which led to the identification of the X-ray afterglow
associated with the GRB. Three further observations were
carried out,
the last performed 6 days after the burst (Fig.1, Tab.1).

LECS(0.1-3.5 keV) and MECS(1.8-10.0 keV) spectra of the observations
were obtained using standard procedures
(BeppoSAX web pages: {\it www.sdc.asi.it}),
with 15-20 counts (source + background) per bin.


Inspection of the light curve (Fig.1) shows that in the first half period
of observation number 1 the flux decays,  followed by a rising
trend in the second half. We have thus   split this 
observation into two parts (1a \& 1b) at the time 
of the minimum  of the light curve, to investigate  
spectral evolution in different states, producing 5 spectra in total.

\subsection{The iron $K_\alpha$ line}

We fitted each  spectrum   
with a power law with variable
absorption, keeping the  relative normalization of LECS vs MECS
free to vary in the range 0.6-0.9. 
This model  provides a good fit to all but dataset 1a.
In this  case we have  $\c2r=2.4$ for $\nu=8$ degrees of freedom, which
corresponds to a probability of
1.4\% \footnote[1]{Note that GB970508 has the highest S/N ratio of the
BeppoSAX GRB afterglows, so that the number of spectra with comparable
statistical weight is limited to those of data set 1. We also remind that
the afterglow of GB970228 had an higher flux, but the S/N ratio was lower,
because data of one MECS only were usable(\cite{Costa97}).}

However the major contribution to $\chi^2$ is not randomly
distributed but is concentrated in 
a line feature around 3.5 keV (Fig. 2a;
note that in figure we show for clarity only data points below 2 KeV for 
LECS and above 2 keV for MECS).

The possible association of this feature with Fe K line emission is obvious,
considering that its  energy coincides with 
\ion{Fe}{1}-\ion{Fe}{20} K$\alpha$ emission at 
the redshift z=0.835, i.e. that of the
putative host galaxy (Bloom et al. 1998). We have therefore added to
the power law model a narrow line
(width $\leq1.5 \times$ FWHM$\simeq 0.5\ $ keV)
at the energy of 3.5 keV.
The relative normalization LECS/MECS -- that depends primarily on the
position of the  source in the LECS detector has, in all datasets,
an error  greater than the nominal range
0.6-0.9.  We have hereafter fixed it to 0.8, the
value averaged over all data sets.
The best fit now yields $\c2r=0.93$ with $\nu=7$. The improvement of  
$\Delta\chi^2/\c2r=13.6$ corresponds to a confidence level of
99.3\% (F test). 
We stress that this confidence level relies on the a priori knowledge of
the energy of the Fe line available from an independent redshift measurement.
Were it not available, the confidence level for the presence of a line 
with free energy would have been 97\%.

The best fit line intensity is $I_{Fe}=(5\pm 2)\times 10^{-5}$
ph cm$^{-2}$s$^{-1}$ (hereafter all errors
and upper limits correspond to 90\% confidence level for a single parameter).
Upper limits on line intensity 
in datasets 1b (Fig.2b), 2 3 and 4 are 2.5, 2, 2.5 and  0.5
 $\times 10^{-5}$ ph cm$^{-2}$s$^{-1}$ respectively. 

The power law parameters obtained in dataset 1a and 1b are consistent
with each other (see Tab. 1) and the absorption column densities are consistent with the hydrogen column in the line of sight of our Galaxy, i.e.
the intrinsic absorption is consistent with zero.
 We next performed a simultaneous fit to both datasets with the same continuum parameters, and the addition of a line to  dataset 1a only.
In this case we left the energy of the line as free parameter, deriving
$E=3.4\pm0.3$, corresponding to
$(6.2\pm0.6)$ in the source rest frame. The error range is too large to
identify the ionization state of iron.
 In fig.2 (inset) we show the contour plot of I$_{Fe}$ vs E.


We note that the absence of the line in dataset 1b explains
why 
a power law model provides an acceptable fit to the
summed spectrum (\cite{Owens98}).
We have also investigated
whether the spectral continuum could be fitted by  a black body 
or a  thermal bremsstrahlung. We find that thermal models  provide
 worse results than that of a power law continuum, as already
shown in the case  of GRB970228 (\cite{F98}).

\subsection{Spectral variability in the afterglow outburst}

Spectral variability of the afterglow spectra supports the
idea that a refreshed energy input produced the 
outburst observed in the light curve.
The spectral index
in the first observation (Tab. 1) is
$\alpha=1.5\pm0.6$. In the second and third observation the flux lies
 above an  extrapolation of the light curve expected from a power
law evolution from the primary event. The spectrum is 
harder ($\alpha=0.4\pm0.6$), indicating electron reacceleration, as
expected in a newly formed shock.  In the last observation the flux
decreases and the spectrum is very steep ($\alpha=2.2\pm0.7$): the new
energy source has been exhausted.
We have noticed previously that spectra 1a and
1b are consistent with each other.  Since the transition 1a-1b 
identifies the beginning of the outburst, one would then expect
an hardening of the spectrum in 1b.
However such an effect could be masked by the continuation of
 the power law decay law (Fig.1), which
accounts for about 60\% of the flux observed in dataset 1b.

\section{Origin of the line and constraints on the emitting region}

The minimum amount of mass $M_{min}$ present can be estimated (Lazzati \etal 
1998) simply by counting how many ionizing photons a given iron atom can
absorb, assuming that recombination is much faster than the time interval
between successive ionizations, and by demanding that the existing iron
atoms thus account for the fluence in the line:
\begin{equation}
\label{min}
M_{min} = 0.1 M_\odot \zeta^{-1}
\left(\frac{I_{Fe}}{10^{-5}\; ph\; s^{-1} \; cm^{-2}}\right)
\frac{T}{10^5\; s} \left(\frac{R}{10^{16}\; cm}\right)^2
\frac{0.1}{q} E_{52}^{-1}
\end{equation}
where $\zeta$ is the iron
abundance in units of the solar value, $T$ is the line duration, 
$R=10^{16} R_{16}\ cm$ is the distance of the medium from the central source,
and $q<0.1$ is
the fraction of the total energy $E=10^{52} E_{52}\ erg$ which is recycled into the line. 
Since the line disappears after about $10^5\; s$, we expect the emitting 
region to have a size $D\approx 3\times 10^{15}\; cm$; in fact, since the
line width is mostly instrumental, we can bar the existence of relativistic 
effects. Then the typical density is $n = 5 \times 10^9 \zeta^{-1} R_{16}^2 \; cm^{-3}$, 
and typical Thompson and iron optical depths are $\tau_T = 10 \zeta^{-1} R_{16}^2$ 
and $\tau_{Fe} = 5 R_{16}^2$. 

>From the above we deduce that the emitting material cannot lie along the line
of sight: for $R_{16} > 1$, $\tau_T \gg 1$, and Thomson scattering
would smear out the time structure of the burst, contrary to observations
(\cite{bdcl98}). 
Even if $\zeta \gg 1$, we would still have $\tau_{Fe} \gg 1$, which implies
an iron edge not seen in the spectrum. For $R_{16} < 1$, we notice that the
fireball would load itself with so many baryons to spoil its relativistic
expansion, again contradicting observations. 

Having established that the emitting medium lies sideways from the
line of sight, 
we also notice that we cannot
have $R \gg D$, because then the line would appear much later than 1 day. 
We conclude that $R \approx D$.  
 
The ionization parameter, using the minimum density derived above, is
found to be
\begin{equation}
\label{Xi}
\xi = \frac{L}{n R^2}= 10^8 \zeta \frac{L}{10^{50}\; erg\; s^{-1}} R_{16}^{-4}
\approx 10^9 \zeta t^{-1.1} R_{16}^{-4}
\end{equation}

where $L$ is the luminosity in the range 0.016-10 keV, whose temporal
evolution is approximately described by a power law (\cite{Piro98}).
>From Eq. (2) 
one sees that for early times ($t \la 10^4\; s$), the material 
will be fully ionized, and thus the line may be due to recombination, while
at later times, when $\xi \la 10^4$, fluorescence may account for the
emission (\cite{kallmann82,hirano87}). 
The exact moment of the transition cannot be determined, because we
established above that the emitting material lies sideways with respect to the
observer, and we cannot exclude some amount of beaming, but the following
discussion still stands. At early times, the emitting material will probably 
reach close to the Inverse--Compton temperature in the burst radiation field, 
which clearly exceeds $10^9\; ^\circ K$; for these high temperatures and low
densities, recombination is not an effective emitter of line photons, and it
is most likely that most of the line we see is emitted by fluorescence
in the time range
$10^4\; s < t < 10^5\; s$, during which the total observed fluence is $\approx 
10^{51}\; erg$. Putting this into Eq. \ref{min}, we find a minimum mass of
\begin{equation}
\label{best}
M_{min} = 0.5 M_\odot \zeta^{-1} \left(\frac{R_{16}}{0.3}\right)^2\;,
\end{equation}
which is our best estimate for the total mass present. 

\subsection{The site of the line emission}

In principle, the line might arise from the normal interstellar medium
surrounding the burst volume, or from a normal stellar wind emitted by
the burst progenitor. However, a very large density ($n\approx 10^9\;
cm^{-3}$) is necessary to 
account for the observed spectral feature; this is obvious when one
considers that the line reported here is as bright as a typical one from 
clusters of galaxies at $z \approx 0.1 \ll z_{970508}$. Winds of massive
stars are also unlikely to be responsible for the iron line: even for
rather extreme wind parameters ($\dot{M} = \dot{m}_{-4} \times 10^{-4} M_\odot\;
yr^{-1}$, and $v = v_2 \times 100\; km\; s^{-1}$), the wind density
exceeds the minimum density only for $R_{16} < 0.1 (\dot{m}_{-4}\zeta/v_2)
^{1/4}$. However, for this small distance, the ionization parameter
(Eq. \ref{Xi}) is $\ga 10^7$ even a day after the burst and, even including 
beaming, the wind 
material will be totally ionized; under these conditions, only recombination
can provide the line photons, but the recombination rates are very small
($t_{rec} > 10^5\; s$), and thus the conditions of instantaneous de--excitation
under which Eq. \ref{min} was established are no longer satisfied: much
larger masses are required than can be accounted for by the winds. Furthermore,
such small distances ($R_{16} < 0.1$) cannot explain the line duration. 
We thus conclude that the material surrounding the burst must have been
pre--ejected. 

$\zeta\approx 1$ is likely to be close to an upper limit to the iron abundance 
of the ISM or of a stellar wind at $z\simeq 1$. Scenarios concerning mergers of 
binaries then begin to appear unlikely, regardless of whether the binary 
is made of  two neutron stars, a black hole and a neutron star, or a black hole 
and a white dwarf, because it has been established (M\`esz\`aros and Rees 1992) 
that they cannot pre--eject, even under the rosiest assumptions, more than 
$\approx 10^{-4}\;
M_\odot$ due to their tidal interactions, about three orders of magnitude
less than the limit in Eq. \ref{min}. We remark that the SupraNova (\cite{vs98})
provides the most appealing scenario, because of the large amount of
iron material ejected by the supernova explosion preceding the
burst. In fact, theoretical computations of the explosion of massive stars
indicate that $\approx 0.1 \Ms$ of iron are ejected by the SN 
(\cite{woosley95,thiel86}). 
In the case of SN1987a, whose progenitor had $M\approx 20 \Ms$,
a similar amount of iron was derived from
optical observations (\cite{danziger}).

In future, it is conceivable
that simultaneous observations of the Ly--$\alpha$ line may help to 
determine $\zeta$, and thus uniquely determine the total illuminated
mass, and by implication the adequacy of different formation scenarios.

\section{Summary}

In this paper we have obtained the following results:
\begin{itemize}
\item
We have searched the X--ray spectrum of GB970508's afterglow for
an iron line, located at the system's redhift (z = 0.835); we found such a
line with limited statistical significance ($99.3\%$) in the early part 
(first $16\; h$) of the afterglow; the line decreases in the later part
of the observations ($\approx 1$ day after the burst) by at least a factor 2, 
enough to make it undetectable with current apparatus;
\item
Simultaneously with the line disappearance, the X--ray flux both
rises and hardens ($\alpha = 0.4\pm 0.6$, while $\alpha = 1.5\pm 0.6$ before 
the reburst), consistent with the appearance of a new shock. Then,
at the
end of the outburst, the spectrum steepens.
\end{itemize}

We showed that these observations indicate that a mass $\approx 0.5  M_\odot 
A_{Fe}^{-1}$ should be located at a distance of $3\times 10^{15}\; cm$
and sideways with respect to the observer,
and is moving at subrelativistic speed. In order to reach such distance, 
this material must have been pre--ejected by the source originating the burst, 
shortly (perhaps a year, for typical SN expansion speeds) before the burst. 
We stress that these observations contain two coincidences: on the one hand,
this is the only burst in which a reburst and a line have been observed
by BeppoSAX; on
the other, the iron line disappears exactly at the moment of the reburst.
We also point out that a line feature, with a similar significance, has been
found by ASCA in another burst, GRB970828, which
also shows an event of rebursting during the X-ray afterglow
(\cite{yoshida98}).

We hope the report of this result
will help shape future observations by X--ray satellites with superior 
observing capabilities, such as XMM, AXAF, ASTRO--E.

\acknowledgements
We thank the referee, Dr. M. Boettcher, for his critical suggestions
that improved the paper substantially, F. Matteucci 
for helpful discussions and the BeppoSAX team for the support with
observations. BeppoSAX is a program of the Italian space agency
(ASI) with the participation of the Dutch space agency (NIVR).

\newpage

\figcaption[fig1.ps]{
The X-ray (2-10 keV) light curve of the X-ray afterglow of GRB980508.  The
dashed line represents the extrapolation from the X-ray data points of
the burst (Piro \etal 1998). For the spectral analysis data from the first
pointing was divided in two parts, corresponding to the 
transition between the
decay and turn up from the light curve, 
creating in total 5 spectra (1a,1b-4).
}

\figcaption[fig2a.ps]{
The spectra (in detector counts) of the afterglow of GRB970508 from data
set 1a (panel a) and data set 1b (panel b). The continuous line
represents the best fit power law continuum in the range 0.1-10 keV,
excluding the interval 3-4 keV.  For clarity we show in figure LECS
data below 2 keV and MECS data above 2 keV.
The inset in panel a) shows the contour plot of 
of the line intensity vs energy.
Contours correspond to 68\%, 90\% \& 99\% confidence levels
for two interesting parameters.
}








\newpage

\begin{deluxetable}{llllllllll}
\footnotesize
\tablecolumns{10}
\tablewidth{0pt}
\tablecaption{Log of Observations and fit with a power law}
\tablehead{
\colhead{Obs.}    &  \colhead{$T_{start}$\tablenotemark{a}} 
&\multicolumn{2}{c}{Exp.Time (ksec)} & \colhead{} &
\multicolumn{2}{c}{Cnt Rate ($10^{-3}$cts s$^{-1}$)} 
& \colhead{$\alpha$} 
& \colhead{$N_H$} & \colhead{$\chi^2/\nu$} \nl
\cline{3-4} \cline{6-7} \\
\colhead{} & \colhead{}   & \colhead{LECS}    & \colhead{MECS} & 
\colhead{}    & \colhead{LECS(0.1-2 keV)}   & \colhead{MECS(2-10 keV)} & 
\colhead{}  & \colhead{($10^{22}$ cm$^{-2}$)} & \colhead{}
} 
\startdata
1a\tablenotemark{b}\ & 0.24 & 8 & 11 & & $3.0\pm0.8$ & $8.8\pm1.2$ 
& $1.5\pm0.9$ & $0.6_{-0.55}^{+1.3}$ & 6.5/7
\nl
1b& 0.65 & 7 & 16 & & $1.4\pm0.7$ & $7.2\pm0.9$ 
& $1.7\pm1.0$ & $1.3_{-1.25}^{+2.5}$ & 6/9
\nl
1a+b\tablenotemark{b} & 0.24 & 15 & 28 & & $2.1\pm0.5$ & $8.3\pm0.8$ 
& $1.5\pm0.6$ & $0.7_{-0.65}^{+1.5}$ & 14.2/18
\nl
2 & 2.74 & 6 & 24 & & $1.3\pm0.7$ & $5.7\pm0.7$ 
& $0.4\pm0.6$ & $0.05^{+1.5}$ &6.2/9 
\nl
3 & 4.14 & 3 & 12 & & $<1.5$ & $3.1\pm0.9$ 
& $0.5\pm0.8$ & $0.05\ fix$  & 5.2/6 
\nl
4 & 5.7 & 14 & 73 & & $1.7\pm0.5$ & $1.2\pm0.3$ 
& $2.2\pm0.7$ & $0.05^{+0.5}$  &6.3/10  
\nl
\enddata
\tablenotetext{}{Errors of spectral parameters correspond to
90\% confidence level for a single parameter of interest;
$N_{HGal}=5\ 10^{20} \rcm2$  }
\tablenotetext{a}{start of the observation in days from the GRB}
\tablenotetext{b}{including a line in spectral fitting of dataset 1a}
\end{deluxetable}



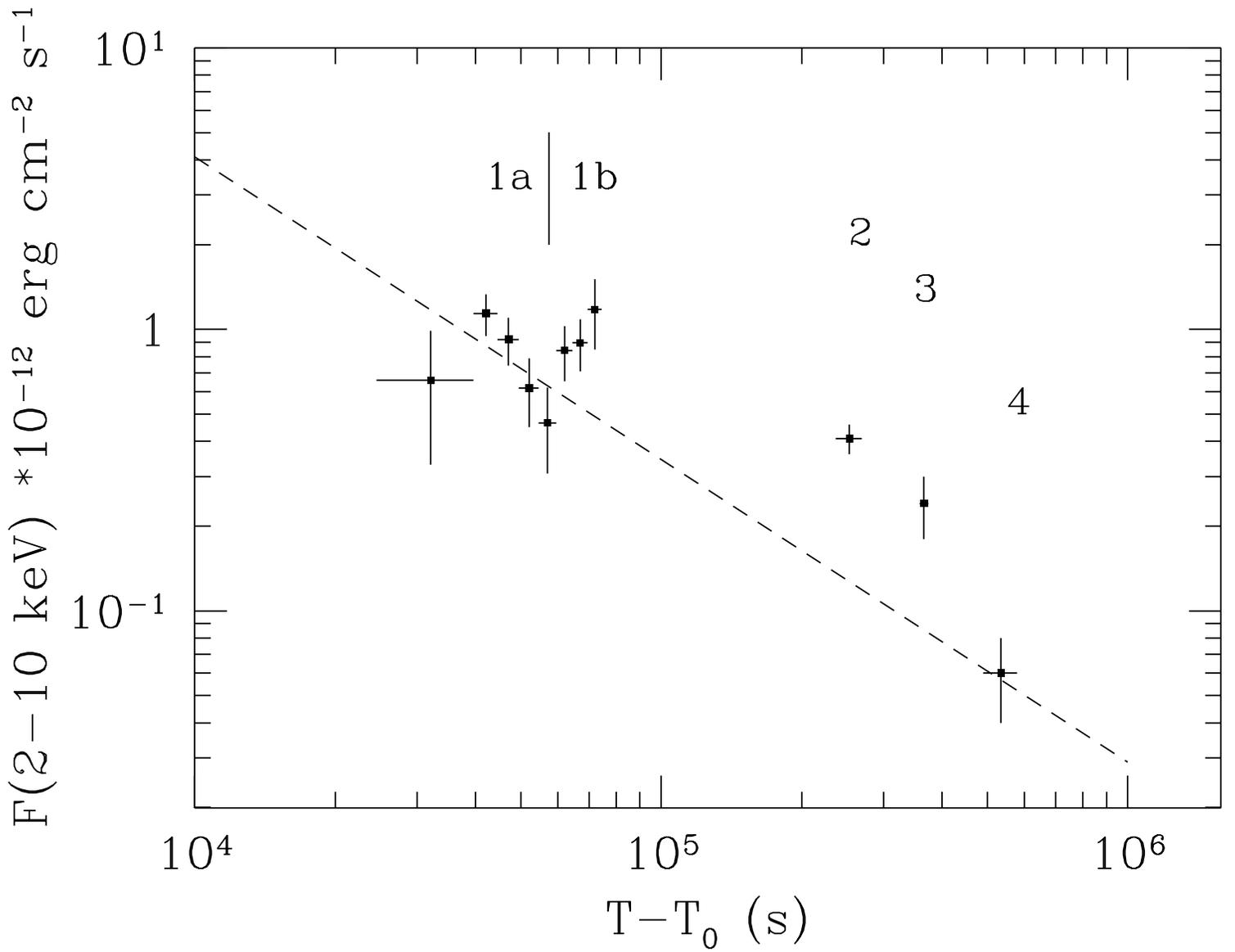
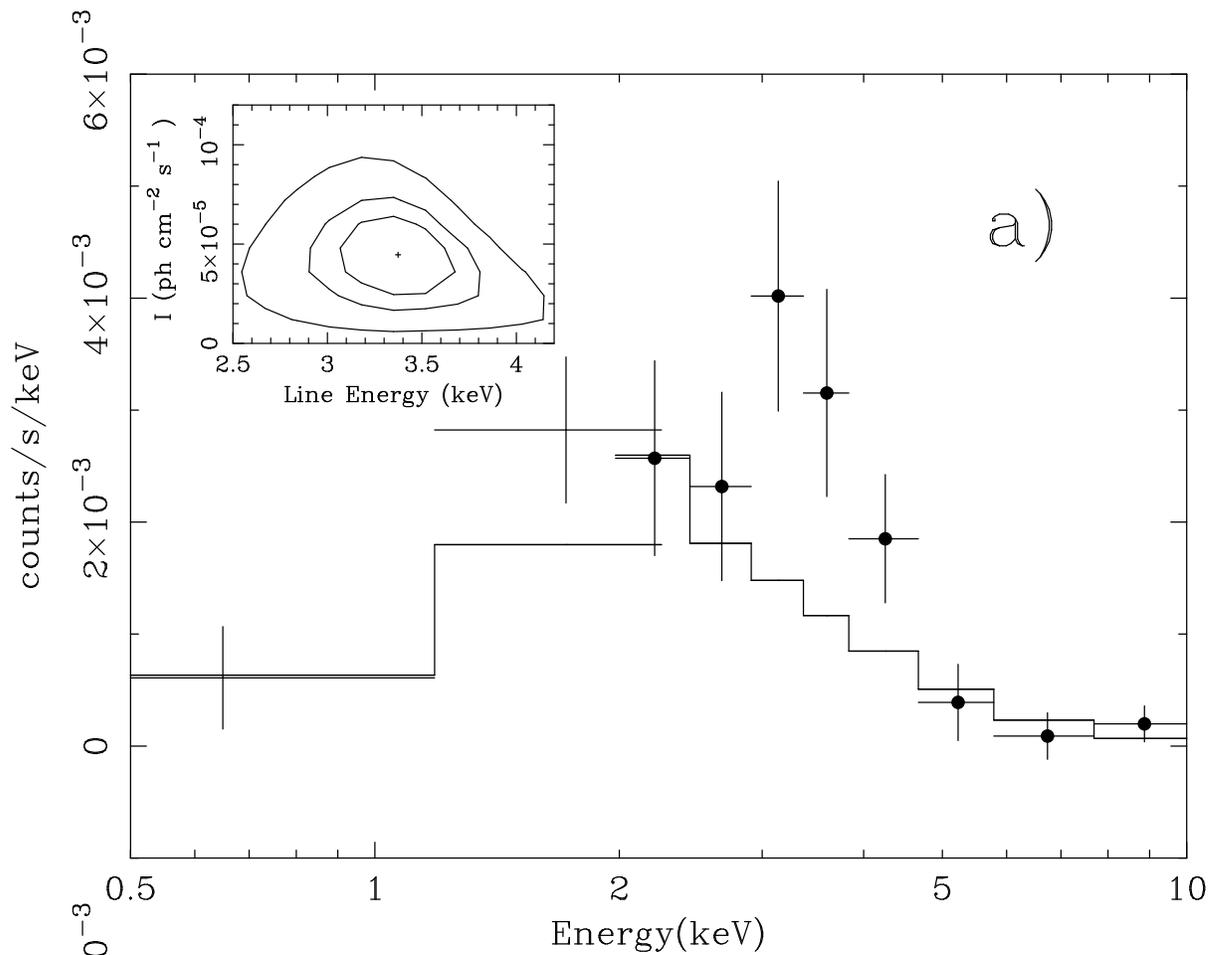
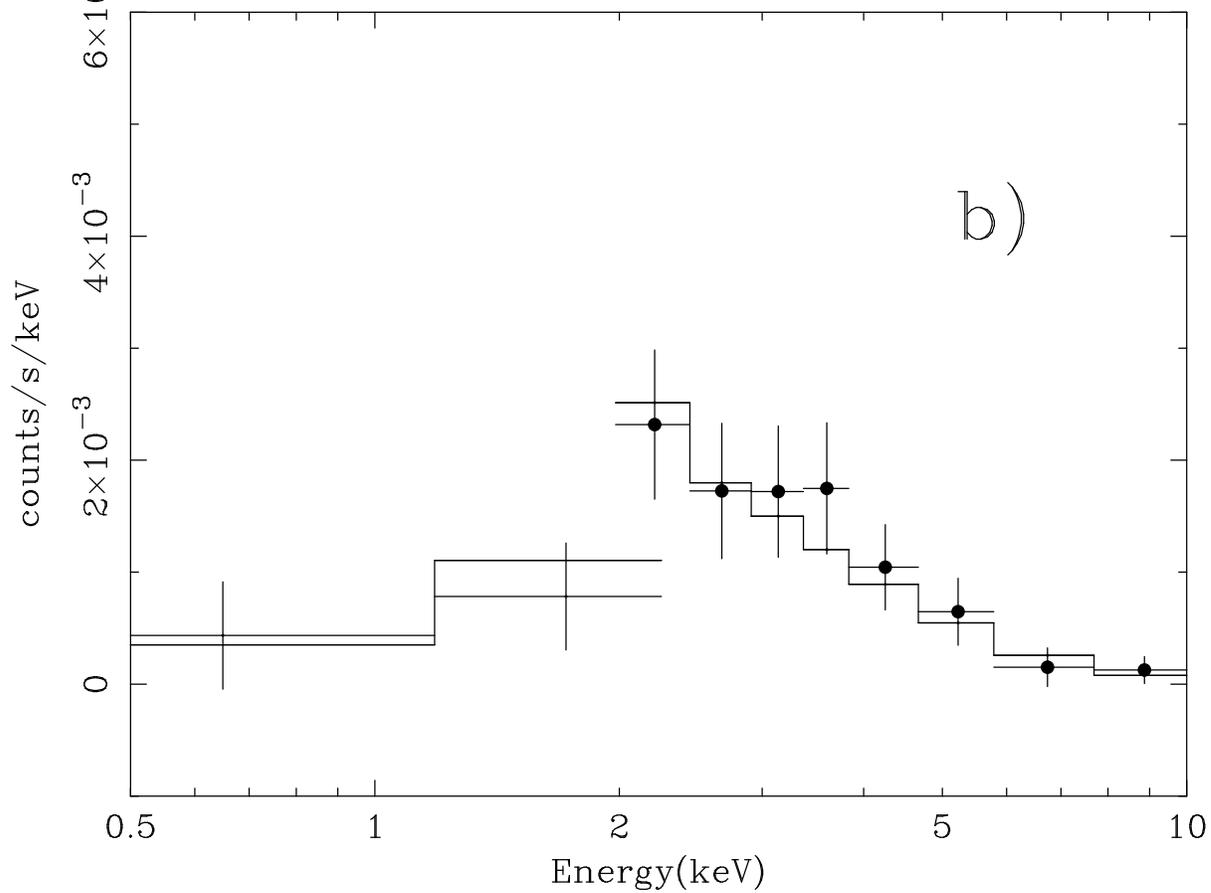
\end{document}